# Application of Machine Learning Techniques for Secure Traffic in NoC-based Manycores

## Progress Seminar - August 2019

Geaninne Lopes, César Marcon (adviser), Fernando Moraes (co-adviser)

*Abstract*—Like most computer systems, a manycore can also be the target of security attacks. It is essential to ensure the security of the NoC since all information travels through its channels, and any interference in the traffic of messages can reflect on the entire chip, causing communication problems. Among the possible attacks on NoC, Denial of Service (DoS) attacks are the most cited in the literature. The state of the art shows a lack of work that can detect such attacks through learning techniques. On the other hand, these techniques are widely explored in computer network security via an Intrusion Detection System (IDS). In this context, the main goal of this document is to present the progress of a work that explores an IDS technique using machine learning and temporal series for detecting DoS attacks in NoC-based manycore systems. To fulfill this goal, it is necessary to extract traffic data from a manycore NoC and execute the learning techniques in the extracted data. However, while low-level platforms offer precision and slow execution, high-level platforms offer higher speed and data incompatible with reality. Therefore, a platform is being developed using the OVP tool, which has a higher level of abstraction. To solve the low precision problem, the developed platform will have its data validated with a low-level platform.

## I. Introduction

NoC-based manycore systems, beyond their inherent scalability, provide massive parallelism and high user performance. The sharing of resources in these systems can be explored through malicious entities, leading to security problems, such as DoS attacks. According to [1], malicious software is responsible for 80% of embedded systems' security incidents. However, in the last decade, some cases of Hardware Trojans also could have resulted in catastrophes. For example, in 2008, a backdoor was found hidden in a microprocessor Sirian radar, which should have alerted the military that an Israel attack was coming [2]. Some attacks come from both Software and Hardware.

The systematic literature review performed in this work presents the different methods proposed in the literature for security in NoCs, such as firewalls, encryption, secure zones, and traffic profiles, among others. Traffic profile methods can detect attacks through network information. For example, studies such as [3] and [4] are within this classification because they use latency information to detect DoS attacks. However, latency can be altered by several factors, such as the execution of multiple applications, leading to a false positive. Therefore, there is a lack of work in learning the actual traffic of the NoC to detect security attacks.

In this context, the main objective of this work is to present a machine learning technique using clustering and time series to detect DoS attacks in manycore NoC. This technique must be validated on a manycore simulation platform, and it is well known that cycle-accurate platforms such as gem5 [5] and hemps [6] can take a long time to execute such algorithms. Conversely, more abstract, instruction-accurate platforms offer faster execution. However, the existing abstract platforms do not support mechanisms such as security control since the data obtained does not match the real ones. Therefore, the second objective of this work is to develop an instruction-accurate platform using OVP and validate its validity by comparing traffic with an RTL platform.

## II. Basic Concepts

This Section presents the Basic Concepts that guided the development of this work. Subsections A and B present NoC and Security concepts, respectively. Subsection C introduces IDS. Finally, Subsections D and E contextualize subsequence time series clustering and probability distribution, respectively.

### A. Network-on-Chip (NoC)

NoC [7] [8] is composed of a set of routers and point-to-point communication channels that interconnect Intellectual Property (IP) components. These components are dedicated hardware modules usually built for reuse in an Integrated Circuit (IC). Figure 1 exemplifies a mesh NoC architecture encompassing IPs, routers, and communication channels (wires) [9] [10]. The routers are composed of a switching structure, control logic for routing and arbitration, and buffers for communication with other components. The communication channels connect two points in the network: a router or an IP. According to [11], using NoCs has advantages, such as greater scalability, bandwidth, and speed, over using buses to interconnect in SoC architectures [12]. Some characteristics that define a NoC are topology, routing, switching, flow control, and arbitration. The following paragraphs describe these characteristics.

NoC topology defines the physical organization of the connections and routers, which can improve the area, energy, and performance costs. Mesh, Figure 2 (a), is a widely used topology consisting of $m$ columns and $n$ rows, commonly referenced as $x$ and $y$ axes, respectively. The routers intersect with two wires, and the computational resources are near routers [14] [15]. Figure 2 (b) exemplifies the Torus topology, an alternative version of the mesh topology since the ends of the columns and lines are connected, providing a greater diversity of routes. Figure 2 (c) exemplifies Fat-Tree topology, which has the computational resources, represented as leaves, and routers, represented as square nodes, organized hierarchically.

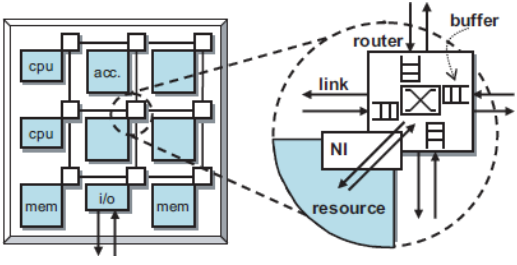

Fig. 1. Example of NoC and Router architectures [13].

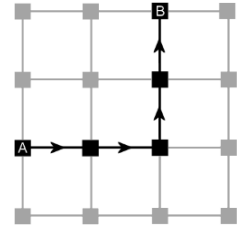

Fig. 3. Example of $XY$ the routing algorithm in a 4x4 mesh NoC. Source: [14]

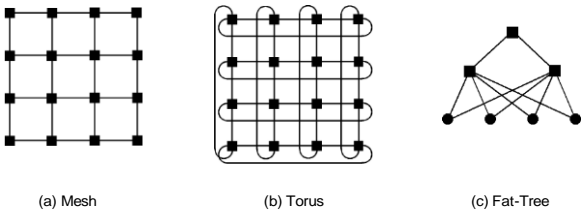

Fig. 2. Example of three NoC topologies [14].

Some arbitration algorithms are Round-Robin, Time-To-Live, and Dynamic Priority. While arbitration grants access to a particular input port, routing algorithms specify the packet's output port. $XY$ is the most used routing algorithm, which deterministically routes packets from the transmitter to the receiver, first in $x$- or horizontal direction to the target column and then in $y$- or vertical direction to the target row, reaching the receiver address [14]. Figure 3 shows the $XY$ algorithm, where node $A$ is the transmitter, and node $B$ is the receiver. Turn model algorithms (West-First, North-last, and Negative-First) are also widely used.

These algorithms must prevent adversities, such as deadlock and livelock. Deadlock may be defined as a cyclic dependency among nodes requiring access to a set of resources so that no forward progress can be made [16]. In livelock, packets keep circulating the network without ever reaching their destination [17]. Starvation happens when a packet in a buffer requests an output channel and is blocked because the output channel is always allocated to another packet [16].

Switching techniques define how data is transmitted in the NoC. Switching techniques can be classified based on network characteristics [18] in two groups: (i) Circuit Switching - a connection is established before data transmission begins, and all path is reserved for that transmission; and (ii) Packet Switching - the message is divided in packets, that are routed individually, from source to the destination. The most used policies in Packet Switching are: (i) Store-and-Forward - packets are entirely stored in the routers before being passed to the next router; (ii) Wormhole - packets are divided into small units called flits, and only the header flit has the routing information, the other flits follow the same path; and (iii) Virtual Cut-Through - packets are forwarded as soon as the next router gives a guarantee that the packet can be wholly stored [19].

### B. Computer System and Network Security

Guttman and Roback [20] define Computer Security as "the protection afforded to an automated information system to attain the applicable objectives of preserving the integrity, avail- ability, and confidentiality of information system resources". Landwehr [21] includes the properties of Non-Repudiation and Authentication. These five principles are described below:

- **Integrity** - Assurance that the information is not modified without proper authorization;
- **Availability** - Assurance that the information is accessible to legitimate users when required;
- **Confidentiality** - Assurance that the information is not disclosed without proper authorization;
- **Authentication** - Assurance that each entity is who it claims to be and
- **Non-Repudiation** - Assurance that the participation of an entity in a transaction cannot be denied.

According to [22], vulnerabilities in a computer system can be divided into seven categories:

- **Architecture / Design vulnerabilities** are consequences of the system structure, such as untested components and centralization of a control system in a single process. If an essential component of the system is centralized in only one location, and there is a failure in this entity, the attacker can exploit it. For example, satellite communication systems often have dedicated terminals, offering a single point at which an attack could cause system failure [22];
- **Behavioral complexity vulnerabilities** are characterized by how the system reacts to variations and whether the reactions are predictable. For example, if the performance of the operating system is sensitive to user load, DoS attacks can be performed by overloading the system;
- **Adaptability and Manipulation**, which encompasses vulnerabilities that stem from the degree to which a

system can be changed by direct user action or induced to change itself in response to such actions [22]. For example, the system can adapt automatically to the attack or to be easily modified by an attacker. On the other hand, if the system is too rigid, it may imply that it cannot be changed in response to an attack.

- **Operation / Configuration changes**, which encompasses vulnerabilities that stem from how a system or process is configured, operated, managed, or administered [22]. For example, systems that have periodic updates and releases in which the security configurations are reset.
- **Non-physical exposure**, which refers to access to devices that do not involve physical contact, such as allowing internet connection, which gives widespread electronic accessibility to a system, allowing the possibility of attacks;
- **Physical exposure**, which refers to vulnerabilities requiring physical access to a device, can be used through electromagnetic radiation to cause damage to hardware equipment, for example, and
- **Dependency on supporting infrastructures**, such as electric power or network connections.

The system security is violated when one of these security properties breaks. Security attacks can cause violations. An attacker exploits a system vulnerability to create unauthorized results through security attacks. According to [23] attacks can be classified as:

- **Passive Attack** - Capture data by monitoring the system without modifying data or changing resources. The passive attacks are divided into two categories: (i) **Eavesdropping** - Information is captured and content read by an attacker; and (ii) **Traffic Analysis** - The attacker can capture information by analyzing the traffic.
- **Active Attacks** - Active attacks are attacks in which the attacker makes changes in the system: (i) **Masquerading** - Occurs when an entity pretends to be another entity; (ii) **Replay** - Data is passively captured and retransmitted for unauthorized effect; (iii) **Message Modification** - Part of the original message is changed to reproduce an unauthorized effect; and (iv) **Denial-of-Service** (DoS) - Occurs when an entity cannot be accessed/utilized.

According to [23], a security mechanism is designed to detect, prevent, or allow recovery from a security attack. For network security, [24] highlights the following mechanisms: Firewall and Intrusion Detection System (IDS). A computer firewall is a software or hardware barrier designed to contain security breaches by monitoring and restricting access to sub-networks [24] [25]. The security provided by the firewall is limited to access restrictions. Despite ensuring that no rules are violated, there are disadvantages to applying this security system in a NoC, such as (i) the more rules, the higher the use of memory, and (ii) it may not work for unknown attacks. IDS aims to detect deviations from the standard behavior of the system, which can indicate an attack.

### C. Intrusion Detection System (IDS)

IDS is an automated defense and security system for monitoring, detecting, and analyzing hostile activities within a network or a host [26]. Figure 4 classifies IDS systems according to the monitored platform and according to the type of detection technique. A network-based IDS analyzes and models traffic, looking for irregular patterns caused by an intrusion. The host-based IDS operates on specific hosts (single PCs) [26], detecting local suspicious activity originating from the target machine or against it. A hybrid IDS considers data provided by host and network, combining the functionalities of both network and host-based IDS [27].

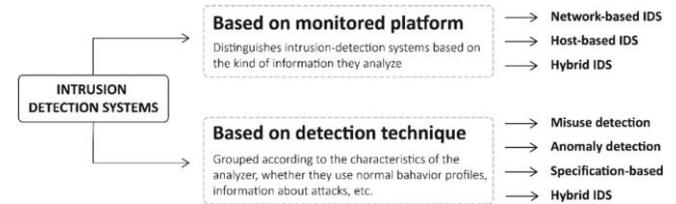

Fig. 4. IDS Categorization. Source: [26]

The detection techniques used in IDS can be grouped as: (i) Misuse detection, which evaluates network activities by using a set of well-known signatures or patterns of attack stored in the IDS database [26]; (ii) Anomaly-based techniques, which are founded on the creation of a baseline profile representing normal network behavior, and any observed deviation of current activity compared to this profile is considered anomalous [26]; (iii) Specification-based, where IDS manually creates specifications and constraints to determine the expected behavior of the network; and (iv) Hybrid IDS, which implements combinations of the already mentioned techniques. Figure 5 presents a general scheme of a network-based IDS. A profile of regular network traffic is generated from the network data in a controlled environment (without attacks). Current network traffic is compared with the generated profile; an attack is detected if the deviation exceeds a pre-defined limit.

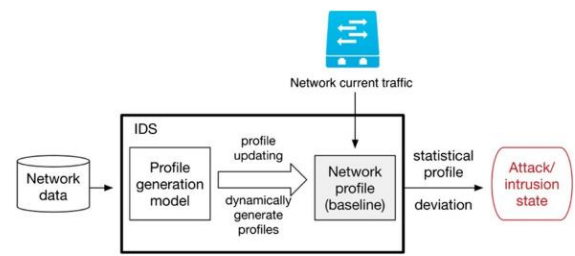

Fig. 5. IDS Scheme. Source: [26]

### D. Sequence Time Series Clustering

A series of data points in similar time spaces is called a time series, usually expressed by line charts [28]. A clustering algorithm can group similar time series sequences in the same cluster. This section presents an overview of clustering methods and distance measures for time series clustering.

*1) **Hierarchical Clustering**:* Hierarchical clustering creates a nested hierarchy of related groups of objects regarding a pairwise distance matrix of the object [28]. Despite having the advantage of not having to pass the cluster number as a parameter, a significant disadvantage of this method is its quadratic complexity. This method can be agglomerative, starting with one cluster per object and grouping until it forms a single cluster, or divisive, starting with a single cluster and dividing into smaller clusters until a pre-defined number of clusters is reached. The agglomerative algorithm is described below, based on [28].

(1) Create a distance matrix with the distance between all objects;

(2) Find the two similar clusters/objects in the matrix;

(3) Join the two clusters/objects to produce a cluster;

(4) Update the matrix calculating the distance between the new cluster and all other clusters and

(5) Repeat steps 2, 3, and 4 until only one cluster remains.

*2) **Partitioning Clustering**:* According to [28], constructs $k$ partitions of the data, where each partition represents a cluster containing at least one object, and $k \leq n$, being $n$ the total number of samples. This method can be divided into fuzzy, where one object can be in more than one cluster, and crisp, where each object belongs precisely to one cluster. K-means and k-medoids are examples of crisp algorithms, and fuzzy c-means and fuzzy c-medoids are examples of fuzzy algorithms. In "means" algorithms, each cluster is represented by the mean value of the objects in the cluster, and in "medoids" algorithms, each cluster is represented by the most centrally located object in the cluster.

Although the k-means algorithm is simple and has a linear computational complexity, it is sensitive to prototype initialization (local minimums) and outliers. On the other hand, the k- medoid algorithm is less sensitive to outliers but has quadratic complexity. The basis of the k-means algorithm is described below (based on [29]).

(1) Determine a value for $k$;

(2) Initialize the $k$ cluster centers;

(3) Assign all the objects to the nearest cluster center;

(4) Re-estimate the cluster centers;

(5) Exit the algorithm if the clusters remain equivalent to the last iteration of the algorithm; otherwise, return to step 3.

*3) **Distance Measures**:* Time series are compared under many aspects, such as complexity, "shape" of the curve, presence or absence of trends and cycles, and invariance. Therefore, it is vital to understand the difference between distance measurements. Distance functions for temporal series are categorized as follows: (i) Minkowski Family- are generalizations of distance used in Euclidean geometry; (ii) Non-elastic Distances - are those that compare two series synchronously concerning the order of observations; and (iii) Elastic Distances - are those that compare two series through an asynchronous alignment of their observations.

Considering two-time series, $S$ and $Z$, represented by vectors $s$ and $z$, Minkowski distances are defined according to (1). When $p = 2$, the Euclidean distance is obtained, which is the most used distance measure. Manhattan distance equals $p = 1$, equivalent to the sum of the straight lines separating two points in each dimension. The p-value can vary to infinity, allowing for many distance metrics. Since the lower the $p$ value, the more small differences between the samples in the series are considered. All metrics in the Minkowski family are non-elastic and, therefore, require that the compared series have the same number of samples. Another non-elastic metric used in this work is the Kullback-Leibler distance, presented in (2), representing the lost information.

$$d(S, Z) = \sum_{t=1}^{n} (|s[t] - z[t]|^p)^{1/p}$$

(1)

$$d(S, Z) = \sum_{t=1}^{n} s[t] * ln(s[t]/z[t])$$

(2)

The lack of uniformity of the temporal axis is a difficulty that can be overcome with elastic distances, such as DTW [30]. DTW can dynamically compare two time-series, even if the data points are not perfectly synchronized. The first step in DTW is to calculate the distances between the temporal series and obtain the distance matrix. For example, given two series $A = \{x_1, x_2, x_3, ..., x_n\}$ and $B = \{y_1, y_2, y_3, ..., y_m\}$, each distance matrix position can be calculated as shown (3), where $i$ and $j$ are the variables that traverse the vectors in a loop. The *minimumNeighbor* (mN) is the lowest value among neighbors already calculated, as shown in (4). Once the distance matrix is built, the warming path can be found, backtracking from the last to the first position— the backtracking algorithm selects the smallest neighbor value. Horizontal moves represent deletion; the period must be excluded from the sequence. Vertical moves represent insertion, which is the period that must be extended. Meanwhile, diagonal moves represent match; both sequences were the same during this period.

$$dist[i][j] = ||A[i] - B[j]|| + mN$$

(3)

$$mN = min(dist[i-1][j-1], dist[i-1][j], dist[i][j-1])$$ (4)

Since, in the first calculations, the matrix positions do not have neighbors, we add an extra row and column in the matrix, where the position of the first row and first column must be filled with 0s, while the rest of the row and column must be filled with the INFINITE; i.e., a value that represents infinite. To show a practical example, Figure 6) displays two series *MODEL* = $\{0, 2, 4, 6, 9, 12\}$ and *TEST* = $\{0, 0, 0, 2, 4, 6, 9\}$. Note that these two series start with the same format but in distinct times.

(5) represents a distance matrix of these two series.

$$
\begin{bmatrix}
 & E & 0 & 0 & 0 & 2 & 4 & 6 & 9 \\
E & 0 & INF & INF & INF & INF & INF & INF & INF \\
0 & INF & 0 & 0 & 0 & 2 & 6 & 12 & 21 \\
2 & INF & 2 & 2 & 2 & 0 & 2 & 6 & 13 \\
4 & INF & 6 & 6 & 6 & 2 & 0 & 2 & 7 \\
6 & INF & 12 & 12 & 12 & 6 & 2 & 0 & 3 \\
9 & INF & 21 & 21 & 21 & 13 & 7 & 3 & 0 \\
9 & INF & 33 & 33 & 33 & 23 & 15 & 9 & 3
\end{bmatrix}
$$

(5)

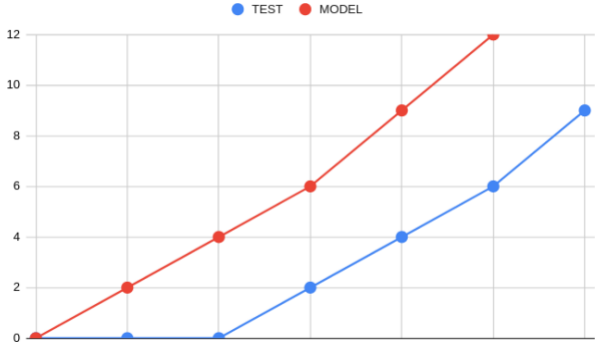

Fig. 6. Example of timing series not aligned.

Once the distance matrix is built, the extra rows and columns can be deleted, and the warming path can be found, backtracking from the last position (dist[n][m]) to the first (dist[0][0]). The backtracking algorithm has to check which is the smallest neighbor value, testing the values of the same row and previous column (dist [i] [j-1]), the same column and the previous row (dist [i-1] [j ]), and the previous column and row (dist [i-1] [j-1]). The warming path of (5) is highlighted in (6).

$$
(6) \quad
\begin{bmatrix}
 & \mathbf{0} & \mathbf{0} & \mathbf{0} & \mathbf{2} & \mathbf{4} & \mathbf{6} & \mathbf{9} \\
\mathbf{0} & 0 & 0 & 0 & 2 & 6 & 12 & 21 \\
\mathbf{2} & 2 & 2 & 2 & 0 & 2 & 6 & 13 \\
\mathbf{4} & 6 & 6 & 6 & 2 & 0 & 2 & 7 \\
\mathbf{6} & 12 & 12 & 12 & 6 & 2 & 0 & 3 \\
\mathbf{9} & 21 & 21 & 21 & 13 & 7 & 3 & 0 \\
\mathbf{9} & 33 & 33 & 33 & 23 & 15 & 9 & 3
\end{bmatrix}
$$

Then, the ... $\{ \ldots , 2, 4, 6, 9, 9\}$, which is aligned to the MODEL sequence in the first 5 periods, as shown in Figure 7.

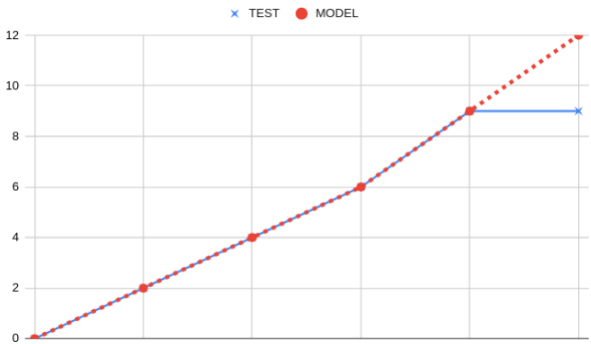

Fig. 7. Two time-series aligned with DTW.

*4) Cluster Evaluation Measures:* There are two main types of group validation indexes: (i) external indexes, which compare the discovered group structure with a previously known group structure, and (ii) internal indexes, which analyze the structure of groups discovered concerning some criteria, such as compactness and separability. In this work, the following external indexes were used: (i) Rand (7); (ii) Jaccard (8); and (iii) Folkes and Mallows (9). Where considering a previously known data clustering G and a clustering T generated by an algorithm, the value of *a* represents the number of pairs that are in the same cluster in G and T, *b* represents the number of pairs that belong to the same cluster in G but do not belong to the same cluster in T, *c* represents the number of pairs that do not belong to the same cluster in G but belong to the same cluster in T, and *d* represents the number of pairs that do not belong to the same cluster in G and does not belong to the same cluster in T.

$$
Rand = \frac{a+d}{a+b+c+d} \tag{7}
$$

$$
Jaccard = \frac{a}{a+b+c} \tag{8}
$$

$$
Folkes and Mallowa = \left( \frac{a}{a+b} * \frac{a}{a+c} \right)^{\frac{1}{2}} \tag{9}
$$

### E. Entropy and Probability Distribution

A probability distribution is a mathematical function that gives the probabilities of occurrence of different possible outcomes for an experiment [31]. For example, if a variable *x* is used to denote the output of a coin toss (experiment), then the probability distribution for *X* = head will be 0.5. Probability distributions can be divided into two classes: (i) discrete probability distribution, where the set of possible outcomes is discrete, and (ii) continuous probability, where possible outcomes can take on values in a continuous range. In this case, probabilities are typically described by a probability density function [32].

The entropy of a random variable is the average level of "surprise" or "uncertainty" inherent in the possible results of the variable. The greater the entropy of a variable, the greater the complexity required for it to be represented. Entropy can be obtained through a histogram, which measures the frequency of values in a data group. The standard methods of calculating entropy are the entropies of Shannon, Tsallis, and Renyi. Shannon entropy [33], which uses the frequency of values and a histogram, is represented by the (10), where $P(X_i)$ is the occurrence probability of $X_i$, *n* is the number of possible values that can be represented, and *B* is the number of possible values for one unity of information. Tsallis entropy [34] is a generalization of Boltzmann − Gibbs entropy [35], which allows the description of non-extensive properties such as fractals and non-linear dynamic systems. Tsallis entropy is represented in (11), where *Q* is the degree of non-extensibility of the domain; the lower the Q value, the more each member of the histogram is considered to have a similar weight within the entropy. Renyi's entropy is similar to Tsallis's and is represented in (12).

$$
S(x) = - \sum_{i=1}^{n} P(X_i) log_B P(X_i) \tag{10}
$$

$$
S(x) = - \sum_{i=1}^{n} \frac{P(X_i)^Q}{Q-1} \tag{11}
$$

$$S(x) = -\frac{1}{Q-1}log_B(\sum_{i=1}^{n} P(X_i)^Q) \qquad (12)$$

## III. OVP BACKGROUND

OVP [36] is an organization created and maintained by Imperas enterprise that promotes open virtual platforms for modeling embedded systems. The set of tools provided by OVP has three main components: (i) **Open Source Models** of several models of processors, memories, and peripherals; (ii) **Simulator and Tools**, such as *OVPSim* and *iGen*, respectively. The first works as an Instruction Accurate Simulator that executes platforms and models coded in OVP APIs, and the second supports the creation of platforms and models; and (iii) **APIs** that operate in the creation and control of platforms, models, and peripherals.

A platform is a collection of components connected into a level of hierarchy in a system to be simulated [36]. The platform consists of a program that makes calls using OP API. Another essential component is a harness file, which is used to instance the platforms and control their execution. A platform can contain open-source models made available by OVP and peripherals or models built by the user. A platform, module, or peripheral model is generated through the **iGen** tool, which reads a TCL input file and outputs a C language file. An example of a TCL file describing a 5-port router is shown in listing 1. In the first line, the command *imodelnewperipheral* indicates that a new peripheral named "router" is being modeled. The following lines define the names of the callbacks that are called when the model is built (constructor) and finished (destructor), respectively. Between lines 9 and 13, the east port of the router is defined and modeled as packetnet, which is a mechanism that provides the interaction between peripherals and other components. As can be seen in Listing 1, it is defined that the data passed by packetnet must have a maximum of 4 bytes (line 11), and *dataEast* callback should be called when there is any writing in this packetnet (line 12).

```
1
2   imodelnewperipheral -name router
3   -constructor constructor
4   -destructor destructor
5   -vendor graph
6   -library peripheral
7   -version 1.0
8
9   # This code must be replicated for each port
10  imodeladdpacketnetport
11  -name portDataEast
12  -maxbytes 4
13  -updatefunction dataEast
14  -updatefunctionargument 0x00
15              . . .
16
```

Listing 1: Example of Router Description in TCL

| Function Name | Description |
|---|---|
| bhmMessage() | Produces a debug output. |
| bhmInstallDiagCB() | Install a callback that is called when the amount of output to be produced, changes. |
| bhmCreateNamedEvent() | Creates an Event. |
| bhmWaitEvent() | Waits an Event. |
| bhmGetSystemEvent() | Returns an Event. |
| bhmTriggerEvent() | Trigger an Event. |
| bhmWaitDelay() | Wait for some time. |

TABLE I
**BHM FUNCTIONS USED IN THE PLATFORM DEVELOPMENT.**

| Function Name | Description |
|---|---|
| PPM PACKETNET CB() | Called when a packetnet is received. |
| PPM REG WRITE CB() | Called when a register will be written. |
| PPM REG READ CB() | Called when a register will be read. |
| ppmPacketnetWrite() | Transmit data to a packetnet port. |
| ppmWriteNet() | Transmit data to a network. |
| ppmOpenAddressSpace() | Open an address Space to be read or write. |
| ppmWriteAddressSpace() | Writes in an address space. |
| ppmReadAddressSpace() | Reads from an address space. |

TABLE II
**PPM FUNCTIONS USED IN THE PLATFORM DEVELOPMENT.**

After defining the peripheral, the TCL file can be passed as input to the iGen tool, which will generate the following five files as output: (i) **User Stubs File**, which contains stub functions, created for each callback, that the user will complete; (ii) **Main File**, which contains the model constructor that will connect buses and netports, and create and initialize registers; (iii) **Include File**, which contains functions and structures declarations, as well as other code required by the C files; (iv) **Attributes File**, which contains the structure that will be interrogated by the simulator when the model is loaded [36]; and (v) **Macros File**, which contains C macros defining register offsets and bit fields [36].

In OVP, each peripheral model runs in its virtual machine, called PSE, with a processor and memory separated from the platform components. This PSE runs the Main file, and the simulator initiates threads and callbacks in the PSE. The code in the model relinquishes control back to the simulator by returning from a callback or by calling a function in the API [36]. Code for the PSE is written using an API split into two parts: BeHavioral Modeling (BHM) and Peripheral Programming Model (PPM). BHM API provides general behavioral modeling capabilities [36]. While PPM API provides an interface to the platform that instances the model [36]. Tables 1 and 2 present the functions used in this work, provided by BHM and PPM API, respectively.

The time model included in the OVP is made by counting instructions the processor executes. The execution time is calculated by dividing the number of instructions executed in all simulation steps by the nominal processor speed in MIPS (Million Instructions per Second). Therefore, OVP is an accurate instruction simulator. The simulation step is called a quantum, which is the number of instructions that each processor executes in each turn [36]; the number of instructions per quantum can be specified by the user in the harness program. The concept of Quantum in OVP can be seen in Figure 7; first, the $P0$ processor executes $x$ instructions, being $x$ the number of instructions per quantum, followed by the execution of $x$ instructions for $P1$, and so on, until all processors have executed $x$ instructions each. The simulation time is advanced only after that, and the next quantum is executed.

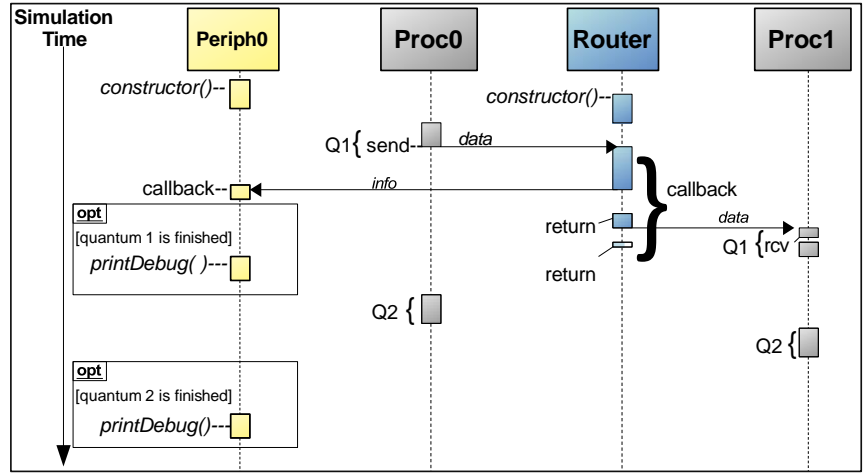

Fig. 8. Example of peripheral and processor executions in OVP.

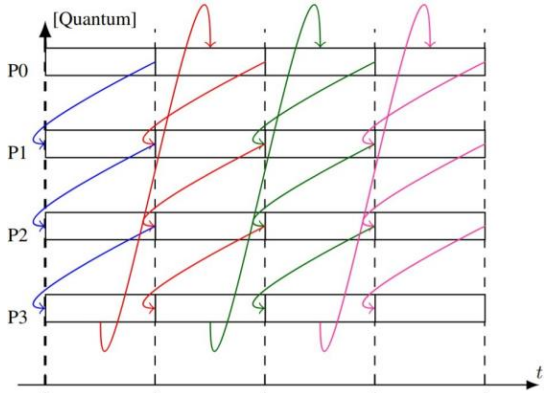

Fig. 9. Quantum concept in OVP - the arrows show the processors' execution sequences.

The quantum only covers instructions executed on the platform's processors, not covering instructions executed on peripherals. The execution of peripherals, in turn, occurs in two ways. First, even before the processors run, the main file of each peripheral will execute. Functions such as bhmWaitDelay () or bhmWaitEvent () are usually inserted in the main file, which serves for the peripheral to execute a specific code from time to time or continuously from a particular event. Second, the peripheral will also perform atomically, from calls to callbacks, which occur when a write or read is on a port. Figure 6 exemplifies the execution of processors and peripherals on the platform, presenting a sequence diagram containing two peripherals (Periph0 and Router) and two processors (proc0 and proc1). The down arrow indicates the simulation time.

First, the peripherals are initialized by installing and connecting ports and registers through the constructor function that is inserted in the main file. This initialization is sequential; the peripheral "Periph0" initializes, and then the "Router". Then, the quantum of the processors starts, with proc0 running first, in this example proc0 sends the message "data" to "proc1" by writing the data in a packetnet that links proc0 with the Router. Then, the callback of this packetnet is called, where the data is sent to its destination "proc1", and the router also sends a message to Periph0 indicating that a transmission was performed. Then, in its turn to execute in quantum 1, "proc1" receives this data. After the completion of quantum 1, "Periph0" calls the printDebug() function that shows the number of transmissions in that quantum on the screen. The peripheral knows that the quantum has run out due to the function bhmWaitDelay (QUANTUM_DELAY), which must be inserted inside an infinite loop in the main file.

Therefore, despite the OVP providing a quantum that simulates the parallelization of processors through instructions. The same is not valid for peripherals; OVP does not provide parallel communication for developing a manycore based on NoC.

## IV. PLATFORM DEVELOPMENT

This Section presents the entire NoC implementation process for the created platform. The first developed version allowed us to understand what would be necessary for NoC development. The second version is more complex, with mechanisms allowing traffic competition and sending messages correctly. However, this version counts with a lack of determinism. Finally, the third version has the advantages of the second version and greater determinism.

*1) First Version:* In the first developed version of NoC, no other mechanism was used to simulate the routers. Thus, the communication works only through the callbacks defined for each packetnet. For example, the packetnet defined in the TCL file of Listing 1 will derive a callback, defined in the User Stubs file, with the name "dataEast", and it will be called when there is a write on the router's east port. This callback was completed in this version, as shown in Listing 2.

```
PPM_PACKETNET_CB(dataEast) {
    int newFlit = *(unsigned int *)data;
    bufferPush(newFlit,EAST);
    outputPort = XYRouting(myID, dest);
    transmit();
}
```

Listing 2: Callback Example.

As presented in the previous section in OVP, time is modeled differently for peripherals and processors. This results in problems, mainly in the NoC simulation, because it is impossible to simulate traffic competition. Secondly, recursion in the callbacks can occur, causing an error in the simulation. In listing 2, the received flit is first stored in a buffer (function bufferPush), then the routing algorithm (function XYRouting) is executed, and the output port is selected. Note that the arbitration algorithm is unnecessary as there is no traffic competition. Finally, the flit is transmitted (transmit function) to the next router, writing to a packetnet from another router and thus calling another callback. In this way, all callbacks in the path will be called, and the open space in the memory will be closed only when the destination processor receives the message. The entire path will be unavailable for new transmissions until it is the target processor's turn to run in quantum.

Figure 10 presents two transmission scenarios, where the order of execution of PEs in the quantum is the following: PE0, PE1, PE2, PE3, PE4, PE5, PE6, PE7, and PE8. In Figure 8(a), PE0 and PE8 send a packet to PE7, even if PE8 sends the packet earlier than PE0 in quantum; since PE0 runs first in quantum, its packet will be sent first. Then PE7, in its turn in quantum 1, will receive the data, and right after, PE8 will send its packet. However, PE7 will only receive the next quantum. In Figure 8(b), on the other hand, PE6 will not be able to deliver the packet; once the data arrives at router 7, it cannot be transferred to the local port because the callback is already open.

Therefore, the development of this more simplistic version resulted in the following conclusions:

- It is necessary to insert a component that acts as a global clock that activates the transmissions. Thus, when writing to a packetnet, callbacks will be terminated as soon as the flit enters the buffer;
- For the traffic flow to be similar to that of a NoC RTL, it is necessary that the order of messages within the quantum be respected. In other words, in Figure 8 (a), if PE8 has sent the packet at the beginning of quantum one and PE0 has sent the packet at the end of quantum 1, it is necessary, if there is no other interference, that the message sent from PE8 to PE7 be delivered first.

*2) Second Version:* Based on the conclusions obtained in the previous version, a new component, an iterator, was added

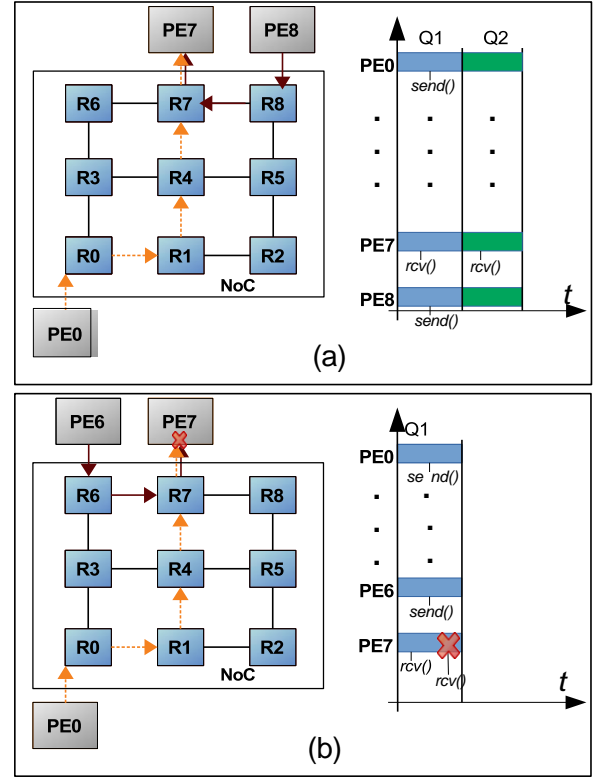

Fig. 10. First Version Scenarios.

to the platform. The iterator activates and orders transmissions and is connected to each router via packet nets, introducing a notion of "coarse-grained time" for the transmissions. In each iteration, each router executes arbitration, routing algorithm, and transmission for one flit.

In addition to informing that new data will be transmitted when a PE makes a new transmission, it will also inform the processor time to the router. This time is obtained through the clock() function. As soon as the router receives this information, it will pass it on to the iterator, stating that a given router has data to be sent and that the processing time for the sending request was requested. Also, a code that runs after each quantum has been introduced to the main file. This code orders the previous quantum requests through the processor time sent by the routers, triggers iterations, and activates transmissions in the correct order.

However, a deeper analysis identified that different scenarios can result in a non-deterministic delivery of packets. For example, depending on the position of the processor in the quantum, it can take 2 or 3 quantums for a packet to be delivered. This brings us to the third and final implementation, which consists of an extra intra-quantum processor that activates the peripheral iterator.

*3) Third and Final Version:* Figure 11 displays the developed simulation platform encompassing the target architecture and its synchronization modules. The target architecture of Figure 11 corresponds to the multiprocessor that will be simulated, which is composed of processing elements (*PE* 0, *PE* 1, *PE* 2, and *PE* 3), and NoC elements - routers (*R*0, *R*1, *R*2, and *R*3) and communication channels (arrows). The synchronization mechanisms are a set of modules that assist the simulation and aim to make the simulation closer to reality. This set has a processing element, called *PENoC*, two peripherals (*iterator* and *sync*), and a bridge connected to each PE (*br*0, *br*1, *br*2, and *br*3).

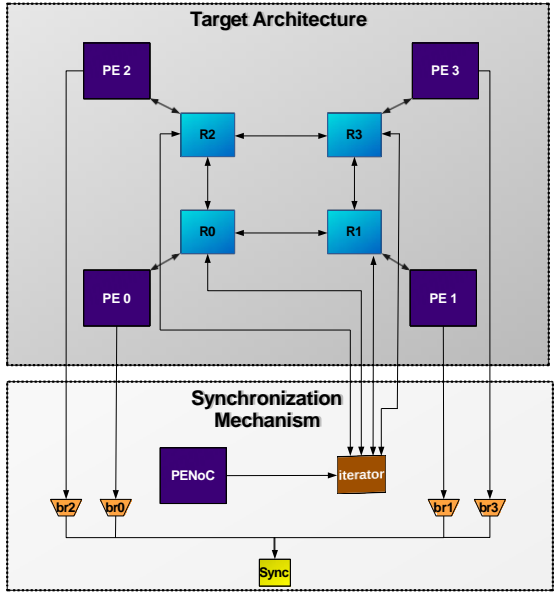

Fig. 11. Scheme of the main components composing the simulation Platform, which is exemplified with four PEs.

Each router has five ports (south, north, west, east, and local) and functions implementing time-to-live arbitration, XY routing, and wormhole flow control. Routers communicate via packetnets, which are mechanisms to notify one component that there is a data structure available to transfer. Then, a pointer is transferred among the routers, indicating the location of the data. The user has to describe application tasks in C, ap- plying an API that provides *send* and *receive* communication functions to use the target architecture. The communication model implemented was based on the Memphis platform, where the transmitter's call of the *send* function only allocates the packet on a waiting list, and the packet is only transmitted when a request comes from the *receive* function on the receptor side.

The synchronization mechanisms help to simulate the target architecture. Below, each module is briefly explained:

- **Sync**: As the processors execute sequentially, the *sync* module synchronizes them at the beginning of the simulation, making the processing only start after all the processors have started;
- **PENoC**: This PE, unlike the others, has only processor and memory since the processor is always the last one to execute within the quantum and aims to make the data transmitted in the NoC;

- **Iterator**: This module also assists the data to be transmitted within the NoC in the correct order and the closest way to the actual traffic.

The implementation architecture is available at https://github.com/iacanaw/OVP_NoC/.

## V. Platform Validation

The platform's validation methodology consists of two phases: visual analysis and comparison and mathematical analysis and comparison. The first phase consists of comparing the developed platform and the RTL-based platform through a Python program. For this purpose, synthetic traffic, such as transposed traffic and hotspots, was created. Figures 15 and 16 show an example of a visualization generated by the Python program. These Figures generated transposed traffic in a 6x6 NoC in the base platform (left) and the OVP platform (right). Each blue square represents a router, and the squares around each router represent its ports and the number of flits that pass through them. These Figures were generated for every 10000 clocks in the base platform and each quantum of 40000 instructions in the OVP platform. It was possible to validate synthetic traffic visually. The next step is to compare traffic from real applications through this visual analysis. Afterward, a comparative analysis will be performed to estimate how many instructions per quantum correspond to a given number of clocks.

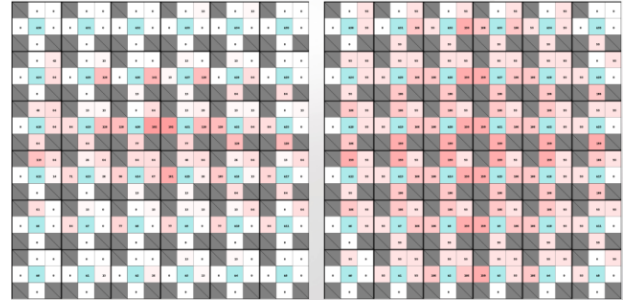

Fig. 12. Transposed Traffic Comparison.

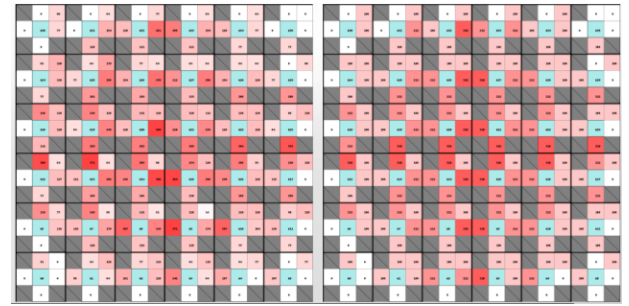

Fig. 13. Transposed Traffic Comparison 2.

## VI. Systematic Review of Techniques for Security in NoCs

We performed a systematic literature review (SLR) to explore the NoC security techniques proposed in academia.

According to [37], SLR is a methodologically rigorous review of research results that identifies relevant studies and synthesizes the research area. The protocol for this review has already been presented in the research project. Therefore, this document presents only the results obtained from this review. Cryptographic schemes can provide the properties of confidentiality, integrity, and authenticity. Such schemes constitute a significant challenge in hardware since they can require high costs of NoC resources, such as performance and area costs. In total, eight studies were found proposing the use of cryptography for security in NoCs. The works propose a series of combinations of cryptographic techniques, such as data encryption [38], [39], [40], [41], [42], Message Authentication Code (MAC) [43], [44], [38], [39] and watermarking [45], that provide confidentiality, integrity, and authenticity, respectively. Additionally, some works can provide authenticity using asymmetric data encryption keys and inserting the packet source in the MAC.

An Error Correction Code (ECC) encodes the messages so that they can be recovered even if some bits are incorrect [46], [47]. This method is used with obfuscation methods in [48], [49], [50], and [51], and with traffic profile in [52]. Obfuscation methods consist of hiding certain information from the attacker. The selected works using this type of method proposed to obfuscate packets [43] [49] [51], links [48], nodes [43] and states in a NI [53]. Packet obfuscation is usually provided through shuffling, scrambling, and permutation methods. Meanwhile, task migration and dummy states provide node obfuscation, and NI states obfuscation.

Routing algorithms select the output port, determining the packet route. The works [54], [50], and [55], and [56] propose routing algorithms targeting DoS and timing channel attacks, where the attacker initiates malicious traffic so that it collides with sensitive traffic and thus obtains some private information. Arbitration algorithms grant the route to specific packets. As proposed in the routing algorithm works, these works [57] and [55] propose countermeasures against timing channel and DoS attacks through specialized arbitration.

A Secure Zone (SZ) is a common mechanism used to protect communication and computation resources [58], isolating an application and the communication among its tasks. Several techniques are used to provide this isolation: routing [59] [60], cryptography [61], mapping [62] and specialized algorithms [58], [63], [64], [65]. Most of these works aim to avoid DoS timing channel attacks.

Firewall solutions work from a set of rules to determine the operations that can be performed. They aim to filter the traffic, block unauthorized data traffic, and release authorized accesses. The first security solution for NoCs was a firewall proposed by [66]. In almost all works [66], [67], [68], [69], [70], [71], and [72] there is a lookup table within the proper access, and the size of this table is proportional to the NoC size, so it is not a scalable solution, and can cause large-area overheads. This is probably a possible cause for this technique to have fallen into disuse.

Regarding Traffic Profile methods, [3] proposed a module within NI, called RLAN, that mitigates attacks against the NoC availability reduction. All packets are tagged with a timestamp, which is used to calculate the latency of a packet. RLAN creates a second packet with the same priority and hop count as the original and compares the latencies in both packets to detect the attacks. The idea behind this design is that two packets traversing routes with significant overlap within a short time are expected to have comparable latencies. [73] used Machine Learning techniques to detect DoS attacks by HTs. Firstly, the authors explored different supervised and unsupervised Machine Learning techniques to detect anomalies in a hierarchical NoC. After precision, recall, and accuracy evaluations, the authors selected the SVM algorithm to be implemented in a hardware module attached to all NoC routers. Besides that, the authors performed a Pearson correlation and classified the following features as the most significant: (i) source core, (ii) destination core, (iii) path, and (iv) distance (hops). Then, when a packet arrives at the destination router, the features of this packet are sent to the detection module, which executes the previously trained SVM algorithm and returns a value indicating if the packet is anomalous or not. [74] proposed router monitors to detect DoS attacks. During the design time, the communication patterns are defined, and some thresholds, such as the maximum number of packets arriving in each router in a time interval and the mean latency of packets arriving in each core, are established. Then, in runtime, the monitors check if the number of packets has been violated. Once an attack is detected, the router sends a signal to its local core, which tries to find the attacker's source by checking the curve latencies and the congestion status in the other routers. Finally, [4] presented one DoS detection technique and two techniques that help us find the attacker's location. When a packet reaches its destination, secure firmware evaluates its latency based on its timestamp. Whether the latency was higher than expected, one of the two proposed techniques can be adopted: (i) Collision Point Router Detection, which changes the packet structure, inserting the router address where the packet waits the longest time and the number of clocks waited; and (ii) Collision Point Direction Detection that, in addition to the previous method, the routers store the competitors by arbitration on that router and the output port that they compete. Thus, detecting the collision point and reducing the list of attacker suspects is possible.

Targeting mainly the works that use the traffic profile, which is the focus of this work, it can be observed that in [3] and [4], the actual traffic is not used since the extracted data are related to latency, and this can easily be altered by several factors, such as the execution of multiple applications, and not only due to an attack. The work of [73] implemented the SVM algorithm of machine learning in hardware to detect DoS attacks. In addition to the fact that the HT inserted in this work would be easily identified at the design stage, the evaluation did not cover the communication of a manycore. Finally, in [74], the number of packets arriving at each router was used for DoS detection; however, the limit values were defined at design time, and if any task is migrated to another processor, all communication can be affected, and, consequently, the proposed technique would be useless.

## VII. Discussion and Scope Definition

As discussed in the previous section, works based on the traffic profile do not use information from real traffic or are limited, not relearning an application's behavior if necessary. Therefore, to detect attacks, a dynamic technique based on the actual behavior of a NoC must be proposed.

In computer networks, IDSs are proposed to detect intrusions in a network. IDSs that use anomaly detection techniques are based on a profile representing the expected behavior of the network, and any deviation from this profile is considered an anomaly. On the other hand, misuse detection techniques use a set of known signatures or attack patterns. In this work, it was chosen to propose an anomaly detection technique, with the justification that it is easier to know and learn the pattern of known applications than the unknown attacks. There is a multitude of studies that propose IDS for networks, which can be found in surveys like [75], [76], [77], [26], and [78]. These surveys generally also analyze the data used in the IDS techniques. Datasets, such as DARPA and KDD, are generally used. The essential features of these datasets come from header information of IP packets and TCP/UDP segments in the tcpdump files. Unlike computer networks, NoCs do not have transmission protocols, and the packets transmitted are not standardized. This means that different NoCs can store different information in the packet headers, and two different NoCs could hardly use the same dataset. Therefore, the initial proposal of this work is to use only data referring to the number of packets that pass through each router in quantum intervals.

To do this, monitors would first need to be inserted in the NoC routers, which could count the number of packets passing in each quantum. This data could not travel through the NoC to reach the security application; otherwise, it would be part of the count and manipulated once the NoC was attacked. Therefore, a second NoC must be inserted to transmit this security data.

Regarding the choice of the type of anomaly detection method, time series anomaly detection methods were also researched due to the following reasons: (i) computer network IDSs usually use a wide variety of features, which is not the case with this work; and (ii) the data extracted from a router at intervals of time form nothing more than time series.

## VIII. Exploration of Solutions

An IDS using time series was proposed by [79] for anomaly detection on electrocardiograms; we chose to use this technique due to the similarity between an electrocardiogram and repetitive NoC traffic. This method uses a sliding window to split a long time series into subsequences and clustering techniques to categorize this subsequence into clusters. The overview of this IDS is shown in Figure 14, where each component is described below:

- **Input Signal**: Also represented as *x*, is the test signal captured from the monitors of the routers;
- **Shape Dictionary**: A dictionary of shapes constructed in the training phase through the clustering algorithms;

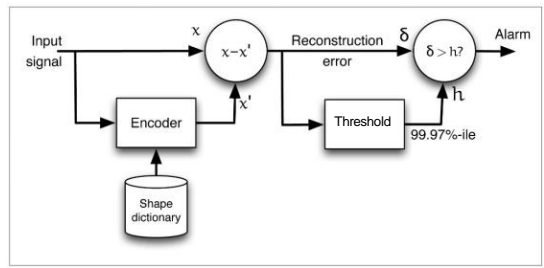

Fig. 14. Window Clustering IDS Method. Adapted from [79].

- **Encoder**: Module that tries to reconstruct the Input Signal, executing the clustering algorithm and reconstructing the signal with the center of the cluster that this signal belongs to;
- **x - x'**: Calculation of the difference between the Input Signal and the reconstructed signal, generating the *Reconstruction Error*;
- **Threshold**: Module that defines the threshold value; and
- **$\delta$ > h**: Verifies if the reconstruction error is bigger than the threshold value;

Figure 15 (a) exemplifies the traffic from a router. As can be seen, some shapes that can be categorized into clusters are repeated over time. In [79], the authors consider electrocardiogram records from hours of execution, and from these records, they created a dictionary of shapes. The example given in Figure 15 is hypothetical; in this work, we also consider records based on many executions for each application. Figure 15 (b) presents the different shapes obtained by the sliding window of the traffic presented in Figure 13 (a). Each shape represents a different cluster. Clustering algorithms can form these clusters.

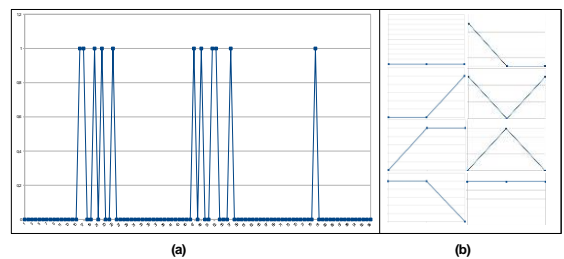

Fig. 15. Example of traffic sequence.

The shape dictionary is used to reconstruct the original input signal by finding the closest cluster to the input signal and using the center value of the cluster to reconstruct the signal. The authors subtract the input signal from the reconstructed signal to construct the error graph. An example is shown in Figure 16, where the graph at the top of the figure shows the input signal, the graph in the middle shows the reconstructed signal and the graph at the bottom represents the error. Whether the error graph surpasses a pre-defined limit value, an anomaly is detected; otherwise, it is considered normal behavior.

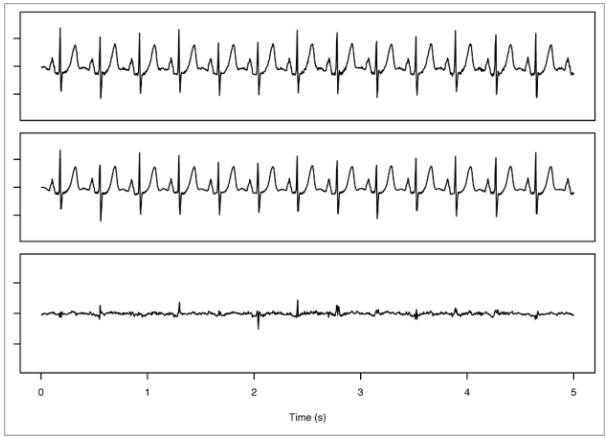

Fig. 16. Reconstruct Error.

This technique was implemented using k-medoids, k-means, and different distance measures. For a first moment, as the platform has not yet been fully validated, a set of shapes, presented in Figure 17 was created. Index Rand, Jaccard, Folkes, and Mallows are used to evaluate the clustering algorithms and distance measures. So, the algorithms that present a better clustering quality for these shapes will be used on the platform. Probability distribution and entropy will be explored in clustering to decrease the memory demand.

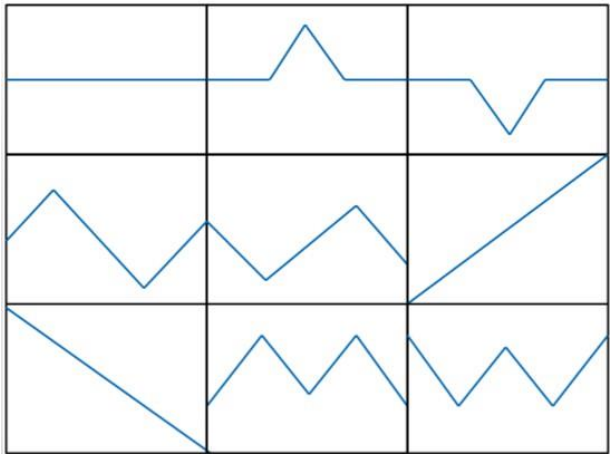

Fig. 17. Set of Shapes.

The platform data will be extracted after validating the platform and choosing the clustering algorithm. This data will be extracted through the monitors contained in each router of the NoC, which count the number of flits that passed through each port and each quantum. Then, these routers will send this data to the proposed IDS at each given quantum range. The data will be generated from a benchmark of parallel applications [80] running on manycore processors and exchanging information via NoC. The idea is to first perform the training in a static mapping and test the technique with the following DoS attacks that are implemented in the related works:

- **Flooding** - Sending intentionally repeated requests to the same destination, wasting bandwidth and causing higher latency transfers in the system;
- **Deadlock** - Changing the route of the packets aiming to disrespect deadlock-free rules; and
- **Misrouting** - Deflecting packets to an invalid address